\documentstyle[12pt,epsf]{article}
\def\thebib#1{
 \list
 {\arabic{enumi}}{\settowidth\labelwidth{[#1]}\leftmargin\labelwidth
 \advance
 \leftmargin\labelsep
 \setlength{\parsep}{0mm}%
 \setlength{\itemsep}{0mm}%
\usecounter{enumi}}
 \def\newblock{\hskip .11em plus .33em minus .07em}
 \sloppy\clubpenalty4000\widowpenalty4000
 \sfcode`\.=1000\relax}

\renewcommand{\thefootnote}{\fnsymbol{footnote}}

\def \invisible{\mbox{$\rule{0mm}{1mm}$}}
\def \mathbox(#1){\invisible\ifmmode{{#1}}\else{\mbox{${#1}$}}\fi}
\def \mbf(#1){\mbox{\boldmath{$#1$}}}
\def\mbfsig{\mbf(\sigma)}
\def\mbfp{\mbf(p)}

\def\calO{{\cal O}}
\def\calA{{\cal A}}
\def\calS{{\cal S}}

\def\calP{{\cal P}}
\def\barcalP{{\overline {\cal P}}}

\def\III{{\rm I\thinspace I\thinspace I}}

\def\Voge{V_{\rm CMI}}

\def\etal{{\it et al.}}

\def\ie{{\it i.e.}}

\def\psiL {\psi_L}
\def\psibR{\psib_R}
\def\psib{\bar\psi}

\def\piii{p_{\rm III}}

\def \mbfp{\mbf(p)}
\def \mbfq{\mbf(q)}

\newcommand{\bra}{\langle}
\newcommand{\half}{\mbox{$\frac{1}{2}$}}
\newcommand{\ket}{\rangle}

\def\VIII{\mathbox(V_{\rm III})}
\def\Voge{\mathbox({V_{\rm OGE}})}

\def\sigsig{\mbfsig_i\cdot\mbfsig_j}
\def\FRAC#1#2{\leavevmode\kern-.em
\raise.5ex\hbox{\the\scriptfont0 #1}\kern-.em
/\kern-.15em\lower.25ex\hbox{\the\scriptfont0 #2}}
\newif\ifrefphysrev

\def\refNP{\refphysrevfalse
           \typeout{** Reference: Nucl Phys format}}
\refphysrevtrue

\def \vol(#1,#2,#3){\ifrefphysrev{{\bf {#1}}, 
{#3} (19{#2})}\else{{{\bf {#1}}(19{#2}){#3}}}\fi}
\def \NP(#1,#2,#3){Nucl.\ Phys.\          \vol(#1,#2,#3)}
\def \PL(#1,#2,#3){Phys.\ Lett.\          \vol(#1,#2,#3)}
\def \PRL(#1,#2,#3){Phys.\ Rev.\ Lett.\   \vol(#1,#2,#3)}
\def \PRp(#1,#2,#3){Phys.\ Rep.\          \vol(#1,#2,#3)}
\def \PR(#1,#2,#3){Phys.\ Rev.\           \vol(#1,#2,#3)}
\def \PTP(#1,#2,#3){Prog.\ Theor.\ Phys.\ \vol(#1,#2,#3)}
\def \ibid(#1,#2,#3){{\it ibid.}\         \vol(#1,#2,#3)}
\makeatletter
\def\scriptsize{\@setsize\scriptsize{14.5pt}\xipt\@xipt
\abovedisplayskip 11\p@ plus3\p@ minus6\p@
\belowdisplayskip \abovedisplayskip
\abovedisplayshortskip  \z@ plus3\p@
\belowdisplayshortskip  6.5\p@ plus3.5\p@ minus3\p@
\def\@listi{\leftmargin\leftmargini
\parsep 4.5\p@ plus2\p@ minus\p@ \itemsep \parsep
\topsep 9\p@ plus3\p@ minus5\p@}}
\def \@magscale#1{ scaled \magstep #1}
\def \half(#1){\mathbox(\frac{#1}{2})}
\def \ninej(#1,#2,#3,#4,#5,#6,#7,#8,#9){\mathbox(\left\{\matrix 
     {#1&#2&#3\cr#4&#5&#6\cr#7&#8&#9\cr}\right\})}
\newif\ifnoncomplete

\noncompletefalse

\def\@cite#1#2{\unskip\nobreak\relax
    {[#1]}} 

\def\citenum#1{{\def\@cite##1##2{##1}\cite{#1}}}
\def\citea#1{\@cite{#1}{}}
 
\newcount\@tempcntc
\def\@citex[#1]#2{\if@filesw\immediate\write\@auxout{%
\string\citation{#2}}\fi
  \@tempcnta\z@\@tempcntb\m@ne\def\@citea{}\@cite{\@for\@citeb:=#2\do
    {\@ifundefined
       {b@\@citeb}{\@citeo\@tempcntb\m@ne\@citea\def\@citea{,}%
{\bf ?}\@warning
       {Citation `\@citeb' on page \thepage \space undefined}}%
{\setbox\z@\hbox{\global\@tempcntc0\csname b@\@citeb\endcsname\relax}%
     \ifnum\@tempcntc=\z@ \@citeo\@tempcntb\m@ne
       \@citea\def\@citea{,}\hbox{\csname b@\@citeb\endcsname}%
     \else
      \advance\@tempcntb\@ne
      \ifnum\@tempcntb=\@tempcntc
      \else\advance\@tempcntb\m@ne\@citeo
      \@tempcnta\@tempcntc\@tempcntb\@tempcntc\fi\fi}}\@citeo}{#1}}
\def\@citeo{\ifnum\@tempcnta>\@tempcntb\else\@citea\def\@citea{,}%
  \ifnum\@tempcnta=\@tempcntb\the\@tempcnta\else
   {\advance\@tempcnta\@ne\ifnum\@tempcnta=\@tempcntb %
\else \def\@citea{--}\fi
    \advance\@tempcnta\m@ne\the\@tempcnta\@citea\the\@tempcntb}\fi\fi}

\def\affiliation#1{\gdef\@affiliation{#1}}
 
\def\and{\cr \makebox[0in]{\rule[-1cm]{0mm}{1cm}and } \cr}

\def\maketitle{\par
 \begingroup
 \def\thefootnote{\fnsymbol{footnote}}
 \def\@makefnmark{\hbox
 to 0pt{$^{\@thefnmark}$\hss}}
 \if@twocolumn
 \twocolumn[\@maketitle]
 \else \newpage
 \global\@topnum\z@ \@maketitle \fi\thispagestyle{plain}\@thanks
 \endgroup
 \setcounter{footnote}{0}
 \let\maketitle\relax
 \let\@maketitle\relax
 \gdef\@thanks{}\gdef\@author{}\gdef\@title{}
 \gdef\@affiliation{} \let\affiliation\relax	%
 \let\thanks\relax}

\def\@maketitle{\newpage
 \null
 \vskip 0em plus 2em minus 0em     
 \ifx\@date\@empty\else
   \begin{flushright}
    {\ifnoncomplete(\today)
     \else{{\normalsize \@date}\\}\fi}      
   \end{flushright}
   \vskip 3em plus 2em minus 2em   
 \fi
 \begin{center}
  {\Large \@title \par}     
  \vskip 3em plus 1em minus 1.5em  
  {
   \lineskip .5em plus 0em minus .3em   
   \begin{tabular}[t]{c}\@author\\
   \end{tabular}\par}
  \vskip 0.5em plus 1em minus 1.5em  
  { \sl \@affiliation \par}
\end{center}
 \par
 \vskip 6em plus 2em minus 4em}     
 
\def\abstract{\if@twocolumn
\section*{Abstract}
\else \normalsize
\fi}
 
\def\endabstract{\if@twocolumn\fi\par\clearpage}

 \makeatother
\refNP
   \input epsf           
\begin{document}
\title{
Spin-Orbit Force of Instanton-Induced Interaction\\
in Strange and Charmed Systems
}
\author{Sachiko Takeuchi}
\affiliation{
Department of Public Health and Environmental Science,\\
School of Medicine,
Tokyo Medical and Dental University,\\
1-5-45  Yushima, Bunkyo, Tokyo 113-8519, Japan}
\date{\today}
\maketitle
\baselineskip=20pt

\noindent
{\bf Abstract:~~}
Effects of the spin orbit-force on hadronic systems with strangeness or 
charm
are investigated by a valence quark model 
with the instanton-induced interaction.
By introducing this interaction, 
the spin-orbit splittings in the negative-parity hyperons 
becomes 0.14 -- 0.37 times smaller.
The flavor-octet baryon mass spectrum and the splittings in the charmed baryons
become consistent with the 
experiments.
Though the splitting is also reduced in the flavor-singlet baryons,
it still gives two third of the experimental value.
The reduction  
comes from the cancellation between the one-gluon exchange 
and the instanton-induced interaction, which is channel-specific.
In most of the two-baryon channels,
the symmetric and antisymmetric spin-orbit force of the YN interaction 
remains strong after introducing this instanton effect.
A few exceptional channels, however, 
are found where the cancellation affects strongly 
and the spin-orbit force becomes small.
\\

\noindent
PACS: 
12.39.Jh, 
13.75.Ev, 
14.20.Jn, 
14.20.Lq  
\\
Keywords:
QCD instanton, spin-orbit force, hyperons, 
hyperon-nucleon interactions, quark model
\\

\newpage
\section{Introduction}
\noindent

Recent experiments on the systems with strangeness are making great progress.
Especially the gamma spectroscopy has identified several gamma transitions, 
which give us valuable information 
on the spin part of the $\Lambda$N interaction
\cite{exp}.
From the observed levels of $\Lambda$-hypernuclei, 
it is believed that the spin-orbit force 
between $\Lambda$ and nucleon is very small 
comparing to that between two nucleons.
It, however, is nontrivial to remove the nuclear effects
to extract the interaction.
Also, only the combined effect of the symmetric and antisymmetric spin-orbit
force can be measured in the hypernuclei.
Information on the noncentral parts of the YN interaction 
has not given directly from experiments yet.
The theoretical investigation has been performed mainly by using the
empirical YN interactions \cite{theoM}.
Here we employ a valence quark model 
to investigate the properties of the spin-orbit force in the strange systems.
The quark model with the instanton-induced interaction, which is introduced 
in this work for the strange systems, 
is found to have an appropriate size and the channel dependence 
for the spin-orbit force and therefore
will enable us to see
 its feature from a more fundamental viewpoint.

Valence quark models
have been applied to low-energy light-quark systems and found to be
successful in 
reproducing major properties of the hadrons and hadronic systems.
These quark models usually 
contains three terms: 
the kinetic term, the confinement term, and the
effective one-gluon exchange (OGE) term 
\cite{MITbag,IK78,Is92,theoNNTokyo,theoYNTokyo,theoYNKyoto}.
It is considered that OGE stands for the
perturbative gluon effects and that the confinement force represents 
the long-range nonperturbative gluon effects.
We argue that a valence quark model 
should include the instanton-induced interaction (\III) as a
short-range nonperturbative gluon effect 
in addition to the other gluon effects
\cite{OT91,TO91,Ta94,Ta96}.

It is well known 
that the color magnetic interaction in OGE 
is responsible to produce many of the hadron properties.
By adjusting the strength of OGE,
the color magnetic interaction can reproduce
the hyperfine splittings (e.g., ground state N-$\Delta$ mass difference) 
as well as the short-range repulsion of
the two-nucleon systems in the relative $S$-wave 
\cite{IK78,Is92,theoNNTokyo}.
It, however, is also known that
the strength of OGE, $\alpha_s$, determined in this
empirical way is much larger than 1, which makes it hard 
to treat it as the perturbative effect.

The QCD instantons were originally introduced in  relation to 
the $U_A$(1) problem.
It produces couplings of instantons to the surrounding 
light-quark zero modes
\cite{tH76,CDGWZ78}.
This leads a flavor-singlet interaction among quarks,
which is believed to be an origin of the
observed large mass difference of $\eta'$-$\eta$ mesons.
How these topological gluonic configurations 
behave in the actual vacuum has not been 
derived directly from QCD.
A few empirical models with such interaction were proposed 
and found to reproduce the $\eta'$ and $\eta$ meson masses and their
properties\cite{OT91,TO95}.
Actually, several recent works 
indicate  that the effects are also large
in the baryon sector \cite{OT91,TO91,Ta94,Ta96,Ko85,MT91,Negele,CNZ96}.
For example, the spin-spin part of 
\III\ produces the nucleon-$\Delta$ mass difference \cite{OT91,Ko85}
and the short-range repulsion between
the two nucleons \cite{OT91} as the color magnetic interaction.  
By introducing \III\, we can reduces the
empirical strength of OGE,  $\alpha_s$ \cite{OT91}.
Moreover, LS of \III\ contributes the spin-orbit force in an interesting way
\cite{Ta94,Ta96}.

The valence quark model including only OGE 
as an origin of the spin-spin interaction
has a spin-orbit problem.
The spin-orbit part of OGE is strong; it is just strong enough 
to explain the observed large spin-orbit force between two nucleons 
\cite{theoNNTokyo}.
On the other hand, 
the experimental mass spectrum of the excited baryons, 
N$^*$ and $\Delta^*$ 
resonances, indicates 
that such a strong spin-orbit force should not exist between quarks \cite{pdg}.
A valence quark model in which the spin-orbit parts 
of the quark 
 interaction are removed by hands 
can well simulate the  observed mass spectrum 
\cite{IK78,Is92,Myhrer,Arima}. 
To explain both of the spin-orbit features at the same time 
is highly nontrivial.

In ref.\ \cite{Ta94}, we demonstrated that 
introducing \III\ may solve the above
difficulty in the nonstrange $P$-wave systems
due to  the channel-specific cancellation between
OGE and \III.
Since here we consider the spin-orbit force on the $P$-wave systems,
the quark pairs which are orbital-antisymmetric and spin-symmetric,
\ie,
only the pairs symmetric (or antisymmetric) simultaneously 
in the flavor and
in the color spaces, are relevant.
It is found that the contribution from the color- and flavor-symmetric
quark pairs dominates in the two-nucleon force,
while only color- and flavor-antisymmetric pairs exist in a baryon.
Because OGE is vector-particle exchange,
and \III\ behaves like scalar-particle exchange,
the sign of their spin-orbit parts is opposite to each other.
Thus, where both of OGE and \III\ contribute,
namely for the color- and flavor-antisymmetric pairs,
LS cancellation occurs.
Therefore, it was expected that introducing \III\ 
would explain the strong LS in the two-nucleon systems
and weak LS in the negative-parity baryons.
In fact, we found such a channel-specific cancellation exists
in the nonstrange $P$-wave systems. 
By introducing 
\III\ strong enough to give the $\eta$-$\eta'$ mass difference,
LS in the single baryons almost vanishes.
The above argument based on the symmetry
 holds even when the system is treated relativistic;
we found the additional $\rho$-space does not
change the conclusion qualitatively by employing a bag model.

We also investigated effects of the noncentral parts 
on the excited nonstrange baryons up to the principal number = 2
and on the two-nucleon systems in a more quantitative way  \cite{Ta96}.
The discussion above is for the relative $P$-wave quark pairs;
there is no such a cancellation between relative even pairs
in the single baryons.
One might wonder 
that the strong LS reappears in the excited positive baryons.
It was found that that LS splittings becomes small
even in the positive-parity baryons because major
 contribution comes from relative $P$-wave pairs even there.
It was also found that LS between the two nucleons remains strong enough
so that more realistic model, a quark cluster model, can actually 
fit the experimental data.

In this paper, we discuss
effects of the spin-orbit part of 
\III\ on the negative-parity hyperons and on the two-baryon systems
with strangeness.

The negative-parity baryon mass spectrum by the present model
shows that there is only weak spin-orbit force between the quarks
due to the above cancellation.
In the present choice of the parameters,
the LS splittings becomes from 0.14 to 0.37 times smaller
than that from the model  only with OGE.
The cancellation also occurs in the flavor-singlet baryons.
There, however, seems to remain enough LS force that 
 will give
an appropriate splitting with the coupling to the N${\overline K}$ channel.

The symmetric and antisymmetric spin-orbit force between two baryons
 remains strong
in most of the channels after introducing \III\ as was found 
in the two-nucleon system,
because the cancellation is weak.
There, however, are a few exceptional channels where OGE-\III\ LS 
cancellation also
gives notable effects:
the symmetric spin-orbit force of the N$\Sigma(I=1/2)$ and 
of the N$\Lambda$-N$\Sigma$
channels almost vanishes after introducing \III.

In section 2, we will explain the model we employed in this paper.
The numerical results and discussions are in section 3.
Summary are in section 4.

\section{Model}

We employed a valence quark model with a non-relativistic 
inter-quark potential.  
This model is chosen because the symmetry is clearly seen, and because
q$^3$ systems can be connected directly to q$^6$ systems,
both of which play essential roles in the discussion in this paper.

\subsection{Interaction}

We introduce \III\ as a short-range nonperturbative gluon effect
in addition to OGE and the confinement potential.
The model hamiltonian for quarks can be written as follows 
\cite{OT91,TO91,Ta94,Ta96,TNK93}:
\begin{eqnarray}
H_{\rm quark} &=& K+(1-\piii)V_{\rm OGE} + \piii \VIII\ + V_{\rm conf}\, ,
\label{eq1}
\end{eqnarray}
where $V_{\rm OGE}$ and \VIII\
are the Galilei invariant terms of the \III\ and OGE potentials.
The parameter $\piii$ 
represents the relative strength of the
spin-spin part of \III\ to OGE.
It corresponds to the rate of the contribution from \III\ to  
the $S$-wave N-$\Delta$
mass difference, $\Delta M_{{\rm N}\Delta}$.
 When one introduces the
interaction strong enough to give the observed $\eta$-$\eta'$ mass 
difference, 408.7 MeV, $\piii$ becomes 0.3--0.4
\cite{OT91,TO91}, which corresponds
90 -- 120 MeV of  $\Delta M_{{\rm N}\Delta}$.  
As we will see later the spectra of the $P$-wave baryons 
including the flavor-singlet $\Lambda$'s 
and those of $\Lambda_c$'s also prefer this size of $\piii$.

First, we derive the nonrelativistic potential for \III.
The effective interaction for \III\ is \cite{OT91,TO91,tH76,Ko85}:
\begin{equation}
 H_{\III} =  \sum_{i<j} V_0^{(2)}(ij) \,\,
   \psibR(i) \psibR(j) \,
\left(
1+{3\over32}\lambda_i\cdot\lambda_j
+{9\over32}\lambda_i\cdot\lambda_j\sigma_i\cdot\sigma_j
\right)
                          \,\psiL(j) \psiL(i)
                          + \hbox{(h.c.)} , \hfill
\label{relaIII}
\end{equation}
where $\psi_{R[L]}={1\pm \gamma \over 2} \psi$,
$\lambda_{i}$ is the Gell-Mann matrix 
of the color SU(3) with
$\lambda_i\cdot\lambda_j = \sum_{a=1}^{8}\lambda_{i}^{a}\lambda_{j}^{a}$,
and $\sigma_{i}$ is the Pauli spin matrix for the $i$-th quark.
$V_0^{(2)}(ij)$ is the strength of the two-body part of \III, 
which survives only when quark $i$ and $j$ have different flavor from 
each other.
The flavor dependence of $V_0^{(2)}(ij)$ can be found from its
relation to the strength of the three-body part of \III, $V_0$,
for the set of quarks with the flavor 
({\it i}, {\it j}, {\it k}) = (u, d, s):
\begin{eqnarray} 
V_0^{(2)}(ij) &=&
{1 \over 2}V_0 \bra {\overline \psi_k}\psi_k \ket \\
&=& {1 \over 2}V_0  {m_k \over m_u}  \bra {\overline u}u \ket\\
&=& \xi_i \xi_j  {1 \over 2}V_0  {m_s \over m_u} \bra {\overline u}u \ket,
\end{eqnarray}
where $m$ is the constituent quark mass, and $\xi_j=m_u/m_j$.
Here, however, we 
take the two-body strength, $V_0^{(2)}(ud)\equiv V_0^{(2)}$ 
as a parameter determined later by eq.~(\ref{eq6}),
and use the flavor dependence of the strength in the above relation.

At the flavor SU(3) limit, eq.\ (\ref{relaIII}) can be 
rewritten as below \cite{OT91}.  
In this form it is more easily seen 
that this interaction operates
only on flavor-antisymmetric quark pairs.
\begin{equation}
 H_{\III} = \sum_{i<j} V_0^{(2)}(ij) \,\,
   \psibR(i) \psibR(j) \,
             {15 \over 8} {\cal A}^{flavor}_{ij}
                ( 1- {1 \over 5} \sigsig )
                          \,\psiL(j) \psiL(i)
                          + \hbox{(h.c.)} , \hfill
\label{relaIIIA}
\end{equation}
where ${\cal A}^{flavor}_{ij} =(1-P^{flavor}_{ij})/2$ 
is the antisymmetrizer in the flavor space. 

We perform the nonrelativistic reduction
to the lowest non-vanishing order in $(p/m)$  
for each operator of different spin structure.
The following potential between quarks of different masses is obtained:
\begin{eqnarray}
V_{\III}&=&V_0(ij)
\left[\left(
1+{3\over32}\lambda_i\cdot\lambda_j
+{9\over32}\lambda_i\cdot\lambda_j\sigma_i\cdot\sigma_j
\right)\right.
\nonumber\\
&+&\left(1+{3\over32}\lambda_i\cdot\lambda_j\right)
\left.\left\{
-{1\over 4 m_i^2}i(\sigma_i\cdot[q\times p_i])
+{1\over 4 m_j^2}i(\sigma_j\cdot[q\times p_j])
\right\}+{9\over32}\lambda_i\cdot\lambda_j \right\{
\nonumber\\
&&
\left.
{1\over 4 m_i^2}i(\sigma_j\cdot[q\times p_i])
-{1\over 4 m_j^2}i(\sigma_i\cdot[q\times p_j])
+{1\over 2 m_im_j}
\left(i(\sigma_i\cdot[q\times p_i])-i(\sigma_j\cdot[q\times p_j])\right)
\right\}
\nonumber\\
&+&\left(1+{3\over32}\lambda_i\cdot\lambda_j\right)\left\{
-{1\over 4m_im_j}(\sigma_i q) (\sigma_j q)\right\}
+{9\over32}\lambda_i\cdot\lambda_j
\left\{{1\over 8}\left({1\over m_i^2}+{1\over m_j^2}\right)(\sigma_i q)
 (\sigma_j q)\right.
\nonumber\\
&&\left.\left.
-{1\over 2 m_i^2}(\sigma_i p_i) (\sigma_j p_i)
-{1\over 2 m_j^2}(\sigma_i p_j) (\sigma_j p_j)
+{1\over  m_im_j}(\sigma_i p_j) (\sigma_j p_i)\right\}\right] .
\end{eqnarray}
Or in a form where the spin-orbit part and antisymmetric spin orbit terms
are gathered separately,
\begin{eqnarray}
V_{\III}
&=&V_0^{(2)} \xi_i\xi_j\left[\left(
1+{3\over32}\lambda_i\cdot\lambda_j
+{9\over32}\lambda_i\cdot\lambda_j\sigma_i\cdot\sigma_j
\right)\right.
\nonumber\\
&+&
\left\{
{9\over32}\lambda_i\cdot\lambda_j
\left({1\over16}(\xi_i^2+\xi_j^2)+{1\over4} \xi_i\xi_j\right)
\right\}
{1\over m_u^2} (\sigma_i+\sigma_j) i [q \times \mbfp_{ij}]
\nonumber\\
&+&\left\{
\left(1+{3\over32}\lambda_i\cdot\lambda_j
 +{9\over32}\lambda_i\cdot\lambda_j\right){1\over16}(\xi_i^2-\xi_j^2)\right\}
{1\over m_u^2} (\sigma_i-\sigma_j) i [q \times \mbfp_{ij}]
\nonumber\\
&+&\left\{
\left(1+{3\over32}\lambda_i\cdot\lambda_j\right)
(-{1\over4}\xi_i\xi_j)+{9\over32}\lambda_i\cdot\lambda_j
{1\over 8}(\xi_i^2+\xi_j^2)\right\}
{1\over m_u^2}\left\{(\sigma_i q) (\sigma_j q)
-{1\over3}(\sigma_i\cdot\sigma_j) q^2\right\}
\nonumber\\
&+&\left.{9\over32}\lambda_i\cdot\lambda_j\left\{
-{1\over 8}(\xi_i^2+\xi_j^2)-{1\over4}\xi_i\xi_j\right\}
{1\over m_u^2}\left\{(\sigma_i \mbfp_{ij}) (\sigma_j 
\mbfp_{ij})-{1\over3}(\sigma_i\cdot\sigma_j) 
\mbfp_{ij}^2\right\}\right] ,
\label{eq:viii2}
\end{eqnarray}
where $\mbfq$ is the three momentum transfer, 
$\mbfp_{ij}=(m_jp_i-m_ip_j)/(m_i+m_j)$.
The first line of eq.~(\ref{eq:viii2}) is the central part, 
the second is the symmetric spin-orbit part, 
the third is the anti-symmetric spin-orbit part, and
the last two lines are the tensor part.
The flavor symmetry is not obvious in eq.~(\ref{eq:viii2}); 
when $m_u=m_d=m_s$, however,
the interaction vanishes for flavor-symmetric pairs 
for each central, spin-orbit, and tensor part.
The same procedure for the one-gluon exchange (OGE) interaction 
 leads \cite{theoYNTokyo,theoYNKyoto,DGG75} 
\begin{eqnarray}
V_{\rm OGE}&=&4\pi \alpha_s {\lambda_i\cdot\lambda_j \over 4} 
{1\over q^2} \left[
1-{1\over8}\left({1\over m_i^2}+{1\over m_j^2}\right)q^2
-{1\over m_im_j} p_i\cdot p_j -
{1\over 4m_im_j}(\sigma_i\cdot\sigma_j)q^2 \right.
\nonumber\\
&+&{1\over 4 m_i^2}i(\sigma_i\cdot[q\times p_i])
-{1\over 4 m_j^2}i(\sigma_j\cdot[q\times p_j])
-{1\over 2 m_im_j}(i(\sigma_i\cdot[q\times p_j])
-i(\sigma_j\cdot[q\times p_i]))
\nonumber\\
&+&\left.
{1\over 4 m_im_j}(\sigma_i q) (\sigma_j q) \right] .
\end{eqnarray}
Or,
\begin{eqnarray}
V_{\rm OGE}
&=&4\pi \alpha_s {\lambda_i\cdot\lambda_j \over 4} {1\over q^2} \left[
1
-{1\over 6 m_u^2}\xi_i\xi_j(\sigma_i\cdot\sigma_j)q^2 \right.
\nonumber\\
&+& {1\over 8 m_u^2}(\xi_i^2+\xi_j^2+4\xi_i\xi_j)
(\sigma_i+\sigma_j)\cdot i [q \times \mbfp_{ij}]
+ {1\over 8 m_u^2}(\xi_i^2-\xi_j^2)(\sigma_i-\sigma_j)
\cdot i [q \times \mbfp_{ij}]
\nonumber\\
&+&\left.
{1\over 4 m_u^2}\xi_i\xi_j\left\{(\sigma_i q) 
(\sigma_j q)-{1\over3}(\sigma_i\cdot\sigma_j) q^2\right\} \right] .
\label{eq:voge2}
\end{eqnarray}
Note that, with $V_0^{(2)}$ negative, spin-spin terms of $V_{\rm OGE}$
and $V_{\III}$ have the same sign while that spin-orbit parts have 
opposite sign.

The central spin-independent part of the interaction $H_{quark}$ 
in eq.\ (\ref{eq1}) has 
the kinetic term, 
the confinement term, in addition to the central parts of $\VIII$ and $\Voge$. 
Once we restrict the 
model wave function to be $(0s)^{n-1}(0p)$ as we will do in this work, 
however,
those central terms only give a constant mass shift for 
the all the states considered here.
Also, we focus our attention to the spin-orbit force and neglect
the tensor part of the interaction.
We will omit these terms in the followings of this paper.

As we mentioned before, \III\ also has a three-body part,
which survives only when it operates on a flavor-singlet set of three quarks.
Moreover, it vanishes when operates on a color-singlet set 
of three quarks; namely,
there is no contribution for the single baryons considered here.
The three-body term may play an important role in two-baryon systems.
In fact, it is reported that
contribution of its central part to the H dihyperon 
is about 50 MeV \cite{TO91}.
Though its spin-orbit part may also contribute to two-baryon systems,
we do not consider the term here for simplicity.

\subsection{Wave functions}

The model wave functions employed here are essentially the same 
as those in the appendix of ref.\ \cite{Ta96}.  
For the single baryons,
the three valence quarks are taken in the color-singlet state, 
the orbital wave function is a harmonic oscillator (0s)$^2$(0p)
with the size parameter $b$, the spin of three quarks is 1/2 or 3/2. 
The base wave function of the flavor part is 
SU(3) singlet, octet, and decuplet, which are mixed 
when the quark mass difference is introduced.
The wave functions in the flavor-octet [21] symmetry in addition to the 
singlet [1$^3$] symmetry 
are summarized in table 1.
 We do not take into account
the orbital wave function deformation caused 
by the quark mass difference nor the
interaction.

As for two-baryon systems, we use
\begin{equation}
\Psi = {\cal A}_{6q}\{ \phi_1 \phi_2 \psi_{0p}(\mbf(R)) \} ,
\end{equation}
where $\phi_i$ is wave function of the $i$-th baryon with a gaussian 
orbital wave function with a size parameter $b$, $\psi_{0p}$ is 
the $0p$ harmonic oscillator wave function of a size 
parameter $\sqrt{2/3} b$, and 
$\mbf(R)=((\mbf(r)_1+\mbf(r)_2+\mbf(r)_3)-(\mbf(r)_4+\mbf(r)_5+\mbf(r)_6))/3$.
Again we do not take into account the deformation of $\psi$. The
matrix element
\begin{equation}
U(R=0) = \bra \Psi | V | \Psi \ket - \sum_i^2 \bra \phi_i | V | \phi_i \ket
\label{eq:adiabaticPot0}
\end{equation}
corresponds to an adiabatic potential for the two baryons at relative
distance, $R=0$, and expresses
the size of the short-range interaction between the baryons.

\subsection{Parameters}

We use an empirical quark model, which contains several parameters
originally.
Their values in the non-strange sector
are the same as in refs.\ \cite{theoNNTokyo,OT91,TO91,Ta94,Ta96}.
The up and down quark mass is 1/3 of the nucleon mass;
the size parameter
$b$ is taken to be a little smaller 
than the real nucleon size reflecting that
the observed baryon size has contribution from the meson cloud;
$\alpha_s$ and $V_0^{(2)}$ 
are determined to give the ground state 
N-$\Delta$ mass difference  $\Delta M_{{\rm N}\Delta}$:
\begin{equation}
{4\over 3\sqrt{2\pi}}{\alpha_s\over m_u^2b^3} 
= -{9\over 4 \sqrt{2\pi}^3}{V_0^{(2)} \over b^3} 
= 293 {\rm MeV} \equiv  \Delta M_{{\rm N}\Delta}~~;
\label{eq6}
\end{equation}
 $\piii$ is taken to be 0.4, which is consistent with 
the strength which gives the $\eta'$-$\eta$ mass difference.
The strange quark mass ratio, $\xi_s = m_u/m_s$, 
is taken to give a correct grand state 
$\Lambda$-$\Sigma$ mass difference.
That for the charm quark, $\xi_c =m_u/m_c$, 
is also determined from the 
$\Lambda_c$-$\Sigma_c$ mass difference.
We use the same $\alpha_s$ for the strange baryons and 
the charmed baryons;
the energy dependence of the OGE strength is neglected.
The results here, including $\piii$ dependence, only depend on above parameters
through $\Delta M_{{\rm N}\Delta}$, $m_ub$, and the quark mass ratio,
whose values are listed in table 2.
The dependence on the values of $m_ub$ and $m_u/m_s$ is not very large, 
so the results do not change much provided that the parameters are taken 
so that eq.~(\ref{eq6}) holds.

When we draw the figure to compare the obtained spectrum to the experiments, 
the kinetic and the central part of the
quark interaction, which give an overall mass shift,
 are substituted by the observed baryon mass average
for each strangeness 0, $-1$ and
$-2$ system. 
They are weighted by spin and charge
degeneracy
over the experimentally established (``four stars") 
octet baryons which corresponds to the
$P$-wave baryons \cite{pdg}: \ie,
N(1535), N(1520), N(1650) and N(1675) for the nonstrange baryons,
$\Lambda$(1670), $\Lambda$(1690), $\Lambda$(1830), $\Sigma$(1670) 
and $\Sigma$(1775)
for the strangeness $-1$ system, 
and $\Xi$(1820), three stars but most established one, 
for the strangeness $-2$ system.

\section{Results and discussions}

\subsection{Single baryons}

Our results of mass spectrum of $P$-wave baryons are summarized in figure 1.
The mass levels expressed by stars (with square blocks 
if the error ranges are given) 
are the experimental values \cite{pdg}.
The levels expressed by horizontal bars are results by the present model 
with $\piii$=0.4: \ie, 40\% of the $S$-wave
N-$\Delta$ mass difference are originated by \III.
We consider that  overall feature of the mass spectrum 
is reproduced quite well by the
present model.  
The spin-orbit force becomes weak due to the same reason as 
the nonstrange negative-parity baryons:
OGE-\III\ LS cancellation in the relative $P$-wave quark pairs.

The most remarkable change in 
the negative-parity single baryon system 
as the flavor space is extended from SU(2) to SU(3) is that
there appear two flavor-singlet states, 
$\Lambda$(1405) and $\Lambda$(1520),
 in addition to other new
octet baryons, $\Lambda^*$, $\Sigma^*$ and $\Xi^*$.

It is a long standing problem whether $\Lambda$(1405), which 
lays below the N$\overline{\rm K}$ threshold, should be considered as
a three-quark state or a bound state of the N$\overline{\rm K}$ system
\cite{pdg,I97}.
Recently, Isgur pointed out that it should be regarded 
as a three-quark state, based on the similarity of its mass spectrum
to the charmed baryons', $\Lambda_c$ \cite{I97}.
This problem should be investigated 
by using a model which
handles the three-quark states and the N$\overline{\rm K}$ system
in a consistent way \cite{Arima,Nogami}.
We, however, use a valence quark model here
and examine what can be said from the valence quark picture.

The problem of the flavor-singlet $\Lambda$ masses 
in the valence quark model prediction can be summarized
into two points:
one is that whether one can produce appropriate LS splitting 
between the two singlet $\Lambda$ particles, the other is 
if there is enough splitting between the singlet and the octet
$\Lambda$ particles.
The former point is directly related to the theme of this paper.
The latter point relates the size of spin-spin interaction, 
the origin of $\Delta M_{{\rm N}\Delta}$.

In our model the LS strength is governed by  $\piii$.
The $\piii$ dependence of $\Lambda^*$ mass spectrum is shown
in figures 2a) and 2b).
The figure 2a) is for the flavor SU(3) limit ($\xi_s=1$),
while the flavor SU(3) is broken 
by the quark mass difference ($\xi_s=0.6$) in figure 2b).
Note that the spin-orbit splittings 
can be much smaller after 
\III\ LS is introduced.
When OGE alone causes the observed $\Delta M_{{\rm N}\Delta}$ ($\piii$=0), 
the calculated mass splittings
between, for example, spin doublet of octet $\Lambda^*$, i.e., 
$\Lambda$(1670) and $\Lambda$(1690), is 224 MeV.
This strong LS 
is canceled by \III\ LS by the same mechanism as the nonstrange baryons.
As $\piii$ increases, the total LS becomes smaller.
At $\piii$=0.5, there is almost no spin-orbit force between quarks.
The present model with $\piii$=0.4 gives 31 MeV for the above mass difference.
The LS splittings in the other channels are also reduced
to from 0.14 to 0.37 times smaller values.
Though the LS-splittings 
seem to be still larger than the experimental ones,
the discrepancy is no more serious, especially 
comparing to the large experimental error ranges.
The dependence of the LS-splittings on $\piii$
is very strong;
when $\piii$=0.42, the above splitting becomes 23 MeV.

The present model with $\piii=0.4$ gives flavor-singlet states
with mass 1470 MeV for $J=1/2$ and 1550 MeV for $J=3/2$.
By changing $\piii=0$ to 0.4, the splitting has become from 333 MeV to 80 MeV; 
while that of the experiment is 115 MeV.
The upper level is 30 MeV higher than the experiment.
Suppose there are other attractive forces 
affecting specifically to the singlet
so that they are 30 MeV more bound, 
then the level corresponding to $\Lambda(1405)$ becomes 1440 MeV; 
which is higher than
the N${\overline {\rm K}}$ threshold only by 8 MeV.
We can conclude that this model with $\piii=0.4$ 
can give bulk amount of the LS-splitting 
between the flavor-singlet baryons and that it can be the pole 
required to explain the low mass of $\Lambda(1405)$ when 
the N${\overline {\rm K}}$ channel is introduced 
\cite{pdg,Arima,Nogami}.

When $m_s=m_u$, the LS-splitting
between $\Lambda$(1405) and $\Lambda$(1520) is proportional to the 
LS-splitting in the $P$-wave octet baryons;
the matrix element is twice larger than
that of the octet baryons,
which holds irrespective of introduction of \III.
That means, if one omits the spin-orbit force between quarks by hands
as the usual way, there is no spin-orbit splittings between the
singlet $\Lambda$'s.

The quark mass dependence of OGE and \III\ is different
from each other.
The mass spectrum with $m_u \neq m_s$ is different 
when one reduces OGE LS by hands
instead of introducing \III.
Suppose the OGE spin-orbit term is weakened to 0.18 times smaller without
introducing \III\ so that the above splitting 
between $\Lambda$(1670) and $\Lambda$(1690)
becomes again 31 MeV, then the splitting of the singlet $\Lambda$'s becomes
54 MeV. The result with OGE-\III\ cancellation, 80 MeV, 
is closer to the experiment, 115 MeV.
Other difference
between these two interaction, which persists at the flavor SU(3) limit,
is in the 
relation between the sizes of the LS splitting of the baryon mass and 
of the LS force of the two baryon systems
and will be discussed in the next subsection.

The mass difference between the nonstrange $S$-wave octet baryons
and the decuplet baryons, 
\ie, $\Delta M_{{\rm N}\Delta}$, is a fitting parameter in our model, 
which determines the combined strength of OGE and \III.
The origin of this mass difference is
the color magnetic interaction of OGE and the spin-spin interaction of \III.
These terms also produce the octet-decuplet mass difference 
in the $S$-wave baryons with strangeness 
and the octet-decuplet and the 
singlet-octet mass difference in the $P$-wave baryons.

The obtained value of the octet-decuplet mass difference of $S$-wave
$\Sigma^*$ particles ($\Sigma$(1385) and $\Sigma^*$) 
is 176 MeV in our model while
the experimental one is 192 MeV.
One can see that the present model 
can give appropriate size and flavor dependence 
for this mass difference.

As for the $P$-wave baryons, 
the above terms divide
the mass levels into three when $m_s = m_u$ 
(see Figure 2a around $\piii=0.5$). 
The highest one consists of the decuplet baryons 
and the octet baryons with the 
quark spin of 3/2.
The middle level consists of the octet baryons with the quark spin of 1/2 .
The lowest one is the singlet baryons.
The splittings between the highest and the middle level is the same as
that between the middle and the lowest level; they are half of
 $\Delta M_{{\rm N}\Delta}$ independent from the value of $\piii$.
Actually the singlet-octet separation here
is much smaller than the observed one.
We do not discuss here the possible origin of the required
additional attraction for the singlet baryons.
Let us mention, however, that the N${\overline {\rm K}}$ channel
has the same flavor symmetry as the ``pentaquark state''
and therefore may have an attractive effect from the quark interaction 
\cite{TNK93} as well as the mesonic effects.
The analysis with coupling to the N${\overline {\rm K}}$ channel
should be required for further consideration.
Also, the observation of the decuplet particles will help to 
see the situation.

\bigskip

Recent experiments indicate that there 
are two ``flavor-singlet'' charmed baryons.
The lower one is $\Lambda_c (2594)$ with spin = 1/2. 
The spin of the upper one, $\Lambda_c (2627)$, is not determined yet;
the decay mode, however, suggests that this state has spin 3/2 \cite{pdg}.
Their splitting is 32.7 MeV.
It is interesting to see if the present model 
gives an appropriate prediction for these two
$\Lambda_c$'s.

The nonrelativity of the model is more 
appropriate for these states; the flavor SU(3), however, is
not valid.
The charmed baryons is considered as systems of  
two light quarks and one charm quark
rather than the flavor octet or singlet states.
The state is more close to the eigenstate of the hamiltonian 
if they are classified
as
\begin{eqnarray}
|\alpha\ket &=& |ud (\makebox{spin=0, isospin=0, relative $S$-wave}), c \ket \\
|\beta \ket &=& |ud (\makebox{spin=1, isospin=0, relative $P$-wave}), c \ket.
\end{eqnarray}
$|\alpha\ket$ corresponds to the total angular momentum, 
$J$ = (1/2), (3/2) states
while $|\beta\ket$ corresponds to $J$ =(1/2)$^2$, (3/2)$^2$, (5/2) 
states.

In figure 2c), 
the  $\piii$ dependence of the mass spectrum 
with $m_c \rightarrow \infty$ is shown.
The solid line corresponds to the $J$=1/2 state. 
The dashed lines correspond to the $J$=1/2 and 3/2 states, 
which are degenerated.
The dot-dashed line correspond to the $J$=3/2 and 5/2 states, 
which are also degenerated.
The mass of the lowest level at $\piii=0.4$ is taken 
to be the observed lowest one, \ie, $\Lambda_c (2594)$.
At $\piii$=0, there is large spin-orbit force between quarks.  
The LS force is canceled
when $\piii$ increases, as the $\Lambda_s$ case.
It vanishes around again $\piii$=0.5, where $|\alpha\ket$ (lower level) and 
$|\beta\ket$ (upper level) are separated only by the spin-spin interaction.

The instanton induced interaction comes 
from the instanton-light-quark coupling. 
Thus, we suppose there is no \III\ 
between the light quarks and the charm quark.
By fitting the observed $S$-wave $\Lambda_c$-$\Sigma_c$ difference, 
168.5 MeV, at
$\piii=0.4$, we obtain
$\xi_c=m_u/m_c=0.23$. It corresponds to $m_c$=1364 MeV;
the value is reasonable considering 
that the current mass of the charm quark is 1.0 -- 1.6 GeV.
The model gives 40 MeV 
for the octet-decuplet mass difference of $S$-wave $\Sigma_c$, 
while the experimental value is 64.5 MeV \cite{pdg}.

In figure 2d), we show the $\piii$ dependence of $\Lambda_c$ mass levels 
at $\xi_c=0.23$.
Again, the mass of the lowest level at $\piii=0.4$ 
is taken to be the observed lowest one.
The overall feature is similar to the fig 2c).
The two lowest levels are
spin 1/2 and 3/2 for $\piii>0.3$ while both 
of them are spin 1/2 for $\piii<0.3$.
The present model at $\piii=0.4$ gives 34MeV for the separation 
between the lowest spin 1/2 and 3/2 states.
The experimental value is 32.7 MeV; the result is remarkably well. 
As seen from the figure, 
this mass difference is insensitive to $\piii$ provided that $\piii>0.3$,
where the second lowest state has spin 3/2.

The wave function of each level gives an information about
whether the strange quark should be treated as a light quark or a heavy quark.
Though it may be too simplified question, 
it helps to see rough properties of the systems.
When $\xi_s=0.6$, 
the flavor-singlet components of the obtained levels corresponding to 
$\Lambda$(1405) and $\Lambda$(1520) are 0.98 and 0.84, respectively,
while the components of $|\alpha\ket$ are 0.63 and 0.85, respectively.
In the lowest flavor singlet state, 
the strange quark acts a light quark.
Its behavior in the higher level is just in-between of light and heavy.
It seems that as far as the mass of the $\Lambda$ is concern, we can treat
 strange quark as a light quark rather than the heavy quark.
The effects of the mass difference in the relative momentum, $\mbfp_{ij}$,
 is included here but it is found to be small and almost indistinguishable 
 in the figure 2b).
Naturally, the charm quark acts more like a heavy quark;
the flavor singlet components of $\Lambda_c$ are  0.79 and 0.42
while the components of $|\alpha\ket$ are 0.89 and 0.94, respectively.

The spin-orbit force, which is our main concern in this paper, 
is a relativistic effect.  
A relativistic model should be used for the check.
As seen in ref.\ \cite{Ta94}, the same conclusion is derived from 
the mass levels of negative-parity N* calculated by the bag model
 as the present non-relativistic model.  
We expect that the situation is similar for the systems with strange quarks.

\subsection{Discussion by the symmetry}

One of the reasons why
valence quark models can be so successful
is that the model space has an appropriate symmetry.
The mechanism of the LS cancellation, 
which affects largely in the excited baryons 
while only at a small extent in the two-nucleon systems, 
is also clearly seen from the discussion on the symmetry 
\cite{Ta94,Ta96}.
Such discussion can also be performed for systems with strangeness; 
which indicates
that the situation in the single baryons and 
most of the strange two-baryon channels is similar to the nonstrange 
systems.
However, there are
a few exceptional YN channels, where \III\ LS affects as strong as OGE LS.

Let us consider the spin-orbit part of the quark interaction
which is color-diagonal and operates on relative $P$-wave quark pairs.
We neglect here the quark mass difference in the relative momentum
because it gives very small effects.
The interaction can be decomposed as:
\begin{eqnarray}
\calO_{\rm qSLS} &=& 
\calO_{\rm qSLS} \barcalP^A + \calO_{\rm qSLS} \calP^S 
\label{eq:projop1}
\\
\calO_{\rm qALS} &=& 
\calO_{\rm qALS} \barcalP^A + 
 \calO_{\rm qALS} \barcalP^S + 
 \calO_{\rm qALS} \calP^S + 
 \calO_{\rm qALS} \calP^A ,
\label{eq:projop2}
\end{eqnarray}
where
\begin{eqnarray}
\barcalP^A  & \equiv & 
\calA^{orb} \calS^{spin} \calA^{color} \calA^{flavor} \\
\calP^S  & \equiv & 
 \calA^{orb} \calS^{spin} \calS^{color} \calS^{flavor} 
\\
\barcalP^S  & \equiv & 
 \calA^{orb} \calA^{spin} \calA^{color} \calS^{flavor}
\\
\calP^A  & \equiv & 
\calA^{orb} \calA^{spin} \calS^{color} \calA^{flavor}  
\end{eqnarray}
with antisymmetrizers $\calA$'s and symmetrizers $\calS$'s.
The operator with bar is for color antisymmetric pairs, 
which is relevant to the single baryons.
Those for the flavor-antisymmetric (-symmetric) quark pairs 
are marked by $A$ ($S$).
The subscript qSLS [qALS] stands 
for the quark spin-orbit interaction which is proportional 
to the $(\sigma \pm \sigma)$ operator.  
$\calO_{\rm qSLS}$ corresponds to the second term in the interaction, 
eq.\ (\ref{eq:viii2})
and the third term in eq.\ (\ref{eq:voge2}), while
$\calO_{\rm qSLS}$ corresponds to the third term in eq.\ (\ref{eq:viii2})
and the forth term in eq.\ (\ref{eq:voge2}).

The matrix elements of OGE LS and \III\ LS 
by a color-antisymmetric and a symmetric quark pair 
are listed in table 3 for each qSLS or qALS term.
AS [S] stands for the contribution from a color-antisymmetric [symmetric]
quark pair, which corresponds to the
first [second] term in eq.\ (\ref{eq:projop1}) 
or to the first two [last two] terms in eq.\ (\ref{eq:projop2}).
The matrix elements of qSLS listed in table 3 are for a ud quark pair;
 those for a us or a ds quark pair can be obtained by multiplying 
 the entry by 
$\xi_s$ for OGE and $\xi^2_s$ for \III.
Those for qALS are for a us or a ds quark pair.

First we consider the system at $m_u = m_s$,
where only  $\calO_{\rm qSLS}$, \ie, the ($\sigma + \sigma$) terms, are 
relevant.
The noncentral term of \III\ contains 
only flavor-antisymmetric component,
 $\calO_{\rm qSLS} \barcalP^A$, while
OGE has both of the components in eq.\ (\ref{eq:projop1}).
Since OGE is vector-particle exchange and \III\ 
is alike to scalar-particle exchange,
their spin-orbit term has an opposite sign
where both of OGE and \III\ exist.
Thus, there is a spin-orbit cancellation for color- 
and flavor-antisymmetric quark pairs.
The LS force in a single baryon, which 
can be obtained by multiplying
the entry of the antisymmetric pair in table 3 by the 
number of quark pairs, three, becomes weak.

We have a new type of spin-orbit
force for the two-baryon systems besides the usual LS.
Since isospin of the $s$-quark is zero, 
NY$^3P_1$ and NY$^1P_1$ may have the same isospin; 
namely,
they are mixed by the strong interaction even 
at the flavor SU(3) limit \cite{theoALS}.
The term which causes this mixing 
is called anti-symmetric spin-orbit force (ALS), and is written 
by the baryon coordinates as
\begin{equation}
V_{ALS} = \sum_{i<j} g_{ij} 
(\sigma_i-\sigma_j)\cdot [\mbfq \times \mbf(P)_{ij}]
\end{equation}
with the coupling constant $g_{ij} = -g_{ji}$.

The {\em symmetric} spin-orbit force in the quark interaction, qSLS,
induces this {\em anti-symmetric} spin-orbit force for the baryons, ALS,
in addition to the ordinary symmetric spin-orbit force (SLS). 
ALS derived from
the quark interaction does not vanish at the flavor SU(3) limit
and can be comparably strong to SLS. 

To obtain the spin-orbit force in the two baryon systems,
one needs to know the factors 
in addition to the size of matrix elements for each 
kind of quark pairs.
Coefficients which should be multiplied to the matrix elements 
in table 3
are listed in table 4 for various two-baryon systems.
For example,
OGE contribution from the flavor 
and color-antisymmetric quark pairs in NN $^3P_1$ SLS is
(3/2)(3/75)$\Delta M_{{\rm N}\Delta}$ = 18 [MeV],
while that from the color-symmetric quark pairs in N$\Lambda$ $^3P_1$ SLS
is ($-$3/2)(6+10$\xi_s$)/36$\Delta M_{{\rm N}\Delta}$ = $-$147 [MeV].

The coefficients for various strange systems listed in table 4 
indicates that contribution from color-symmetric quark 
pairs dominates in most of SLS and ALS in the strange two-baryons systems.
In the two-nucleon system, for example,
it can be seen from the fact that the coefficient of S, 38/75, 
is much larger than the coefficient of AS, 3/75.
There, the size of SLS or ALS we obtain 
with non-zero $\piii$
is approximately equals to the one obtained
by reducing the strength of OGE to $(1-\piii)$.
Note that, suppose one reduces the LS splitting of the single baryons 
by weakening OGE LS by hands, the size of LS here can be 
about 20 \% of the full OGE value, 
which is much smaller than the present result, 
about 60 \%.

The exceptions are SLS of
N$\Sigma$($I=1/2$) and N$\Lambda$-N$\Sigma$ channels.
The contribution from the color-antisymmetric quark pairs
are as large as the symmetric one in these channels,
which means that \III\ LS gives large effects.
The OGE-\III\ LS cancellation plays important role there,
 and the LS force becomes also small similarly to the LS splittings
in the single baryons.

The numerical calculation was performed also at $m_u\neq m_s$, 
where $\calO_{\rm qALS}$ as well as the quark mass dependence 
of $\calO_{\rm qSLS}$ 
should be taken into account.
Actually, the factor of qALS in table 4 is not very small comparing to qSLS.
However, 
the size of the matrix element of one pair is about ten times smaller
as seen in table 3.
Thus we expect that the results will change only by order of ($1-\xi_s$)
when the quark mass difference is taken into account, 
which is confirmed by the numerical calculation.

\subsection{Two-baryon systems}

We investigate
$U_{\rm SLS[ALS]}(R=0)$ defined by eq.\ (\ref{eq:adiabaticPot0}), 
the adiabatic potential
between the two baryons
 at zero relative distance.
This value is considered to express the strength of the short-range
interaction between the two baryons.

The  numerical results at $\xi_s$ = 1 and 0.6 
with and without \III\ are listed in table 4.
The size of ALS is comparable to SLS in general.
The size of both kinds of the spin-orbit forces depends strongly 
on the channels.
SLS of N$\Lambda$ and N$\Sigma$(I=3/2) diagonal,
ALS of N$\Sigma$ $^3P_1$-$^1P_1$,
 and ALS of N$\Lambda$ $^1P_1$-N$\Sigma$ $^3P_1$ are
as strong as SLS of NN.
ALS of N$\Sigma$(I=3/2) $^3P_1$-$^1P_1$ is very small.
Size of both kinds of the spin-orbit forces in the other channels 
are from 20\% to 40\% of NN SLS.

When $\xi_s$ changes from 1 to 0.6, 
N$\Lambda$ SLS is reduced by about 25\%
in both of the $\piii$ = 0 and $\piii$ = 0.4 cases.
SLS of N$\Lambda$-N$\Sigma$, ALS of N$\Sigma$ 
$^3P_1$-$^1P_1$,
and ALS of N$\Lambda$ $^1P_1$-N$\Sigma$ $^3P_1$ are reduced by about 10\%
at $\piii$ = 0.
ALS of N$\Lambda$ $^3P_1$-N$\Sigma$ $^1P_1$ at $\piii$ = 0 
is reduced by about 20\%
while ALS of N$\Lambda$ $^3P_1$-$^1P_1$ at $\piii$ = 0.4
increases by about 20 \%.
The spin-orbit force in other channels, however, 
are almost the same by changing $\xi_s$.

Introducing \III\
changes the channel dependence of the spin-orbit force.
It was reported that if the interaction between baryons 
holds the flavor SU(3),
which corresponds to the $\piii$=0 and $\xi_s=1$ case here, 
the channel dependence is determined only by the SU(3) symmetry 
\cite{theoALS}.
The $U_{\rm SLS[ALS]}(R)$ from qSLS at $\piii$=0 and $\xi_s=1$ or 0.6,
were calculated and found to hold 
the above relation between the channels
except for the factor from the norm kernel \cite{theoALS}.
Since \III\ affects color- and flavor-antisymmetric quark pairs selectively, 
this relation  deviates when \III\ is switched on.
As we discussed in the previous subsection, 
there are channels where the contribution from the color-antisymmetric pairs
 is comparably important to 
the symmetric ones. 
In the N$\Sigma$($I=1/2$) and the N$\Lambda$-N$\Sigma$ channels,
 we found that introducing \III\ changes actually 
the relative strength of the spin-orbit force between the baryons 
considerably.

The adiabatic potential at $R>0$ looks like a 
gaussian with the range of about 1 fm \cite{theoYNTokyo,theoALS}.
Since the potential we are considering here
is SLS or ALS between relative $P$-wave, 
the potential at $R>0$ 
will be more important.
Moreover, when we treat the quarks dynamically by, {\it e.g.},
 a quark cluster model,
the potential we should consider between baryons
is not
the adiabatic one but the RGM potential, which is highly
nonlocal \cite{theoNNTokyo,theoYNTokyo,theoYNKyoto}.
We argue, however, as far as a relative strength of SLS or ALS 
 to the NN SLS is concerned, the conclusion here
holds even when
one performs more 
sophisticated calculations.

The ALS term is also in the meson-exchange interaction \cite{theoALS};
it was found that the tensor couplings of the vector-meson exchange 
can produce ALS at the flavor SU(3) limit. 
Its channel dependence is determined by SU(3).
When the SU(3) is broken, various meson exchange produce the ALS term.
The meson-induced ALS seems much smaller than
that of quarks, though the size of the meson coupling is not well known
\cite{theoALS}.

Experimentally, information on the YN spin-orbit force has been given only 
through the level splittings in $\Lambda$ hypernuclei.
The $_\Lambda^9$Be data 
suggest the N$\Lambda$ spin-orbit interaction is very small \cite{exp}.
The level splittings, however, gives only the combined strength
of SLS and ALS; the strong SLS and ALS obtained from 
the quark model may cancels each other.
It was reported that the other effect such as the YN tensor interaction
may reduce the splitting \cite{theoM}.
More investigations both from the theoretical and experimental sides
are necessary 
to understand the spin properties of the systems with strangeness.

\section{Summary}

We investigate the effects of the quark interaction induced 
from the instanton-light-quark coupling
on the spin-orbit force in the negative-parity hyperons mass spectrum 
and in various relative $P$-wave two-baryon systems
with strangeness.
The spin-orbit force of this instanton-induced interaction (\III) affects only 
the color-antisymmetric ones among the relative $P$-wave quark pairs.
It cancels the spin-orbit force from the one-gluon exchange (OGE)
for those pairs.  Thus there is OGE-\III\ LS cancellation 
where the color-antisymmetric pairs play  an important role.

We employ a valence quark model where the strength of \III\ is 
determined
empirically:
 its spin-spin part gives 40\% of the ground state
nucleon-$\Delta$ mass difference.
The value is  consistent with the strength
which can give the
$\eta'$-$\eta$ mass difference.
The negative-parity baryon mass spectrum by the present model 
shows that there is only weak spin-orbit force between the quarks
due to the above cancellation.
In the present choice of the parameter set, the
LS splittings 
are from 0.14 to 0.37
times smaller than that of the model only with OGE.
The splittings in the flavor-octet baryons becomes consistent with 
the experiments.
The splitting between the flavor-singlet $\Lambda$'s 
becomes about two third of the experimental value.
There seems to remain enough LS force that it will 
give an appropriate splitting with the coupling 
to the N${\overline K}$ channel.

Most of the symmetric and the antisymmetric spin-orbit force between 
two baryons remains strong after introducing \III\
as was found in the two-nucleon system, because the color-symmetric
quark pairs play an important role there.
There, however, are a few exceptional channels where OGE-\III\
LS cancellation are also large: as a result,
the symmetric spin-orbit force of the
 N$\Sigma$($I=1/2$) and of the N$\Lambda$-N$\Sigma$ channels become small.

The conclusion will be more quantitative if one extends the model,
{\it e.g.}, to take into account the meson effects or the deformation of the 
relative motion wave functions between the baryons.
The properties of the excited baryons and YN interaction 
should be investigated both in the theoretical and in the experimental way.

\smallskip

The author would like to thank K.\ Yazaki and M.\ Oka 
for valuable discussions.
This work was supported in part by the Grant-in-Aid for scientific research 
Priority Areas (Strangeness Nuclear Physics) of the Ministry of Education, 
Science, Sports and Culture of Japan.

\newpage

\newpage
\noindent
Figure 1: Negative-parity baryon mass spectrum.\\

\noindent
Figure 2: The $\piii$ dependence of the $P$-wave $\Lambda^*$ mass.\\

Each graph corresponds to the case where
a) quark mass ratio, $\xi_{s}=m_u/m_s=1$, b)  $\xi_{s}=0.6$, 
c) no interaction between the charm and other quarks, 
and d)  $\xi_{c}=m_u/m_c=0.23$.\\
In figures a), b) and d), solid lines stand for the spin 1/2 states, 
dotted lines
 for the spin 3/2 states, and dot-dashed lines for the spin 5/2 states.
Spin quantum numbers of the degenerated states are written in figure c).

\newpage
\noindent
Table 1: Single baryon wave functions in the flavor space.\\

The SU(3) wave function in the mixed [anti]symmetric (MS[MA])
of the [21]
symmetry and in the [1$^3$] symmetry are given.
For the [21] states with non-zero $z$-component of the isospin, 
the flavor of $q_1$ and $q_2$ and overall phase ($P$)
of the wave function
$|{\rm MA}\ket = P{1\over\sqrt{2}}(q_1q_2-q_2q_1)q_1$, and
$|{\rm MS}\ket = P{1\over\sqrt{6}}(2q_1q_1q_2-q_1q_2q_1-q_2q_1q_1)$
are listed.
For the other states, coefficients of the $q_1q_2q_3$ components
together with the normalization $n$ are listed.
\\

\noindent
\begin{tabular}{l|ll|r}
\hline
\hline
B & $q_1$ & $q_2$ &  $P$ \\
\hline
\hline
p          & u & d & $+$ \\
n          & d & u & $-$ \\
$\Sigma^+$ & u & s & $-$ \\
$\Sigma^-$ & d & s & $-$ \\
$\Xi^0$    & s & u & $+$ \\
$\Xi^-$    & s & d & $+$ \\
\hline
\end{tabular}
\\

\noindent
\begin{tabular}{ll|rrrrrr|c}
\hline
B          &    &  uds &  dus &  dsu &  usd &  sud &  sdu & $n^{-2}$ \\
\hline
\hline
$\Sigma^0$ & MA &    0 &    0 & $-$1 & $-$1 &    1 &    1 &  4 \\
           & MS & $-$2 & $-$2 &    1 &    1 &    1 &    1 & 12 \\
$\Lambda^8$& MA &    2 & $-$2 & $-$1 &    1 & $-$1 &    1 & 12 \\
           & MS &    0 &    0 & $-$1 &    1 &    1 & $-$1 &  4 \\
$\Lambda^1$&  A &    1 & $-$1 &    1 & $-$1 &    1 & $-$1 &  6 \\
\hline
\hline
\end{tabular}
\bigskip
\bigskip

\noindent
Table 2: Model parameters.\\

\bigskip

\noindent\begin{tabular}{cccc}
\hline
\hline
$\Delta M_{{\rm N}\Delta}$[MeV] &  $m_u b$  & $m_u/m_s$& $m_u/m_c$\\
\hline
293 & 0.79 & 0.6 & 0.23 \\
\hline
\hline
\end{tabular}
\\
\bigskip
\bigskip

\newpage
\noindent
Table 3: The spin-orbit matrix elements of two-baryon systems.\\

Matrix elements of OGE and \III\ 
by each color-antisymmetric (AS) and symmetric (S) quark pairs are
listed in unit of the grand state N-$\Delta$ mass difference, 
$\Delta M_{{\rm N}\Delta}$.
Those from the ($\sigma+\sigma$) term in the quark potential, qSLS, are 
for ud quark pairs; for us or ds pairs, $\xi_s$ for OGE, 
$\xi_s^2$ for \III\ should be
multiplied.
Those from the ($\sigma-\sigma$) term , qALS, are 
for us or ds pairs.
These matrix elements are to be multiplied by the factors listed in table 4.
\\

\noindent
\begin{tabular}{ll|ccc}
\hline
\hline
&& AS & S  \\
\hline
qSLS  &OGE  &           3/2 & $-$3/4\\
& \III\ & $-$1/$m^2b^2$ & 0\\
\hline
qALS&OGE & $(1-\xi_{s}^2)$/4 & $-$$(1-\xi_{s}^2)$/8\\
 & \III\       & 0 & 0\\
\hline
\hline
\end{tabular}
\\
\bigskip
\bigskip

\noindent
Table 4: Symmetric and antisymmetric spin-orbit forces 
in the two-baryon systems.\\

Size of the SLS and ALS forces of various two-baryon systems are shown.
Coefficients which should be multiplied to the entries in Table 3 
are listed for each of the color-antisymmetric
quark pair (AS) and the symmetric pair (S) 
and for each of ud and us quark pair, which 
is to be divided by the entry under $n^{-1}$.
The obtained $U(R=0)$ in eq.\ (\ref{eq:adiabaticPot0}) for each
$m_u/m_s=1$ and 0.6, and the contribution of \III, $\piii$,
is 0 (OGE only) and 0.4 are listed  separately in MeV together
with the relative strength to NN SLS.
\\

\setlength{\tabcolsep}{0.5mm}
\hspace*{-1.5cm}
\noindent
\begin{tabular}{lccc|rrrr|rr|r|rr|rr|rr|rr}
\hline
\hline
&&& & \multicolumn{4}{c|}{qSLS}&\multicolumn{2}{c|}{qALS}&&\multicolumn{8}{c}{$U(R=0)$}\\  
\cline{12-19}
	& &  & &\multicolumn{2}{c}{AS}&\multicolumn{2}{c|}{S}&AS&S& & \multicolumn{4}{c|}{$m_u/m_s=1$} & \multicolumn{4}{c}{$m_u/m_s=0.6$} \\
\cline{12-19}
	& 2$T$ & $S$ & $S'$  &ud&us&ud&us&us&us&$n^{-1}$& \multicolumn{2}{c|}{$\piii=0$} & \multicolumn{2}{c|}{$\piii=0.4$} & 
\multicolumn{2}{c|}{$\piii=0$} & \multicolumn{2}{c}{$\piii=0.4$}\\
\hline
SLS\\
\hline
NN-NN                 & 2 & 1 & 1 &   3&   0&    38&    0&    0&    0&            75& $-$94 &    1.00 & $-$61 &    1.00 & $-$94 &    1.00 & $-$61 &    1.00 \\
N$\Lambda$-N$\Lambda$ & 1 & 1 & 1 &   1&   1&     6&   10&    2&    2&            36& $-$73 &    0.78 & $-$51 &    0.83 & $-$54 &    0.57 & $-$37 &    0.61 \\
N$\Sigma$-N$\Sigma$   & 1 & 1 & 1 &  69&$-$3&    90&$-$10&$-$18&$-$42&           516&    22 & $-$0.24 &  $-$3 &    0.05 &    22 & $-$0.23 &  $-$4 &    0.06 \\
N$\Lambda$-N$\Sigma$  & 1 & 1 & 1 &$-$3&$-$3&     8&    0&    0& $-$4&12$\sqrt{129}$& $-$32 &    0.34 & $-$14 &    0.22 & $-$28 &    0.30 & $-$13 &    0.21 \\
N$\Sigma$-N$\Sigma$   & 3 & 1 & 1 &   3&   3&    69&    7& $-$9& $-$3&           150& $-$94 &    1.00 & $-$61 &    1.00 & $-$96 &    1.02 & $-$61 &    0.99 \\
\hline
ALS\\
\hline
N$\Lambda$-N$\Lambda$ & 1 & 1 & 0 &   1&$-$2&     6& $-$2&    1&   10&            36& $-$37 &    0.39 & $-$18 &    0.30 & $-$37 &    0.40 & $-$23 &    0.38 \\
N$\Sigma$-N$\Sigma$   & 1 & 1 & 0 &   9&  18&$-$102& $-$6&    1&$-$10&36$\sqrt{129}$&    87 & $-$0.93 &    44 & $-$0.71 &    79 & $-$0.85 &    43 & $-$0.69 \\
N$\Lambda$-N$\Sigma$  & 1 & 1 & 0 &   3&$-$6&    14& $-$2&    3&   10&           108& $-$37 &    0.39 & $-$18 &    0.30 & $-$30 &    0.32 & $-$19 &    0.31 \\
N$\Lambda$-N$\Sigma$  & 1 & 0 & 1 &   3&   6& $-$30& $-$6&    3&$-$10&12$\sqrt{129}$&    87 & $-$0.93 &    44 & $-$0.71 &    79 & $-$0.84 &    42 & $-$0.69 \\
N$\Sigma$-N$\Sigma$   & 3 & 1 & 0 &$-$9&   9& $-$15&   15& $-$1& $-$5&  90$\sqrt{5}$&     0 &    0.00 &     0 &    0.00 &  $-$1 &    0.01 &     3 & $-$0.05 \\
\hline
\hline
\end{tabular}

\newpage
\epsfbox{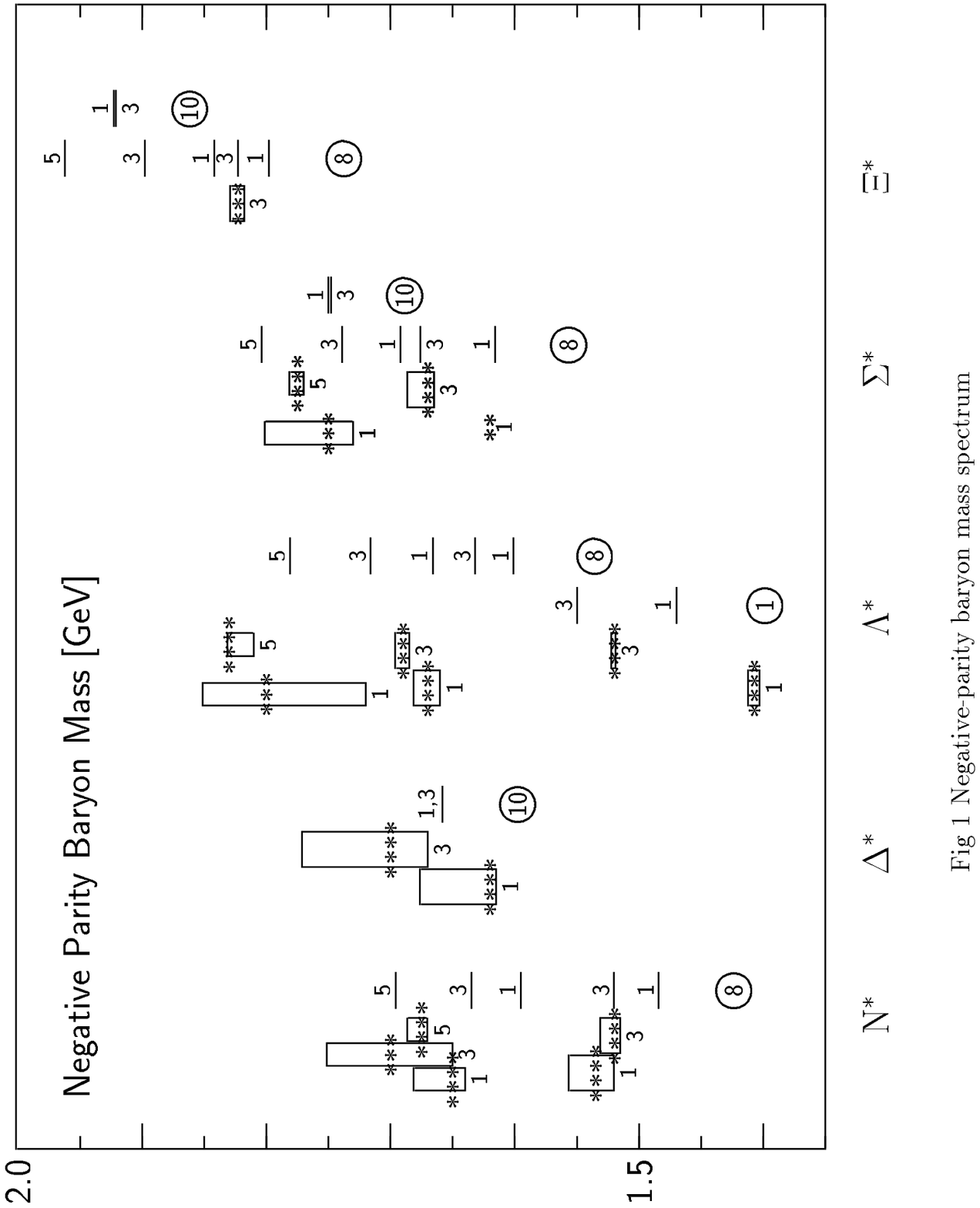}
\epsfbox{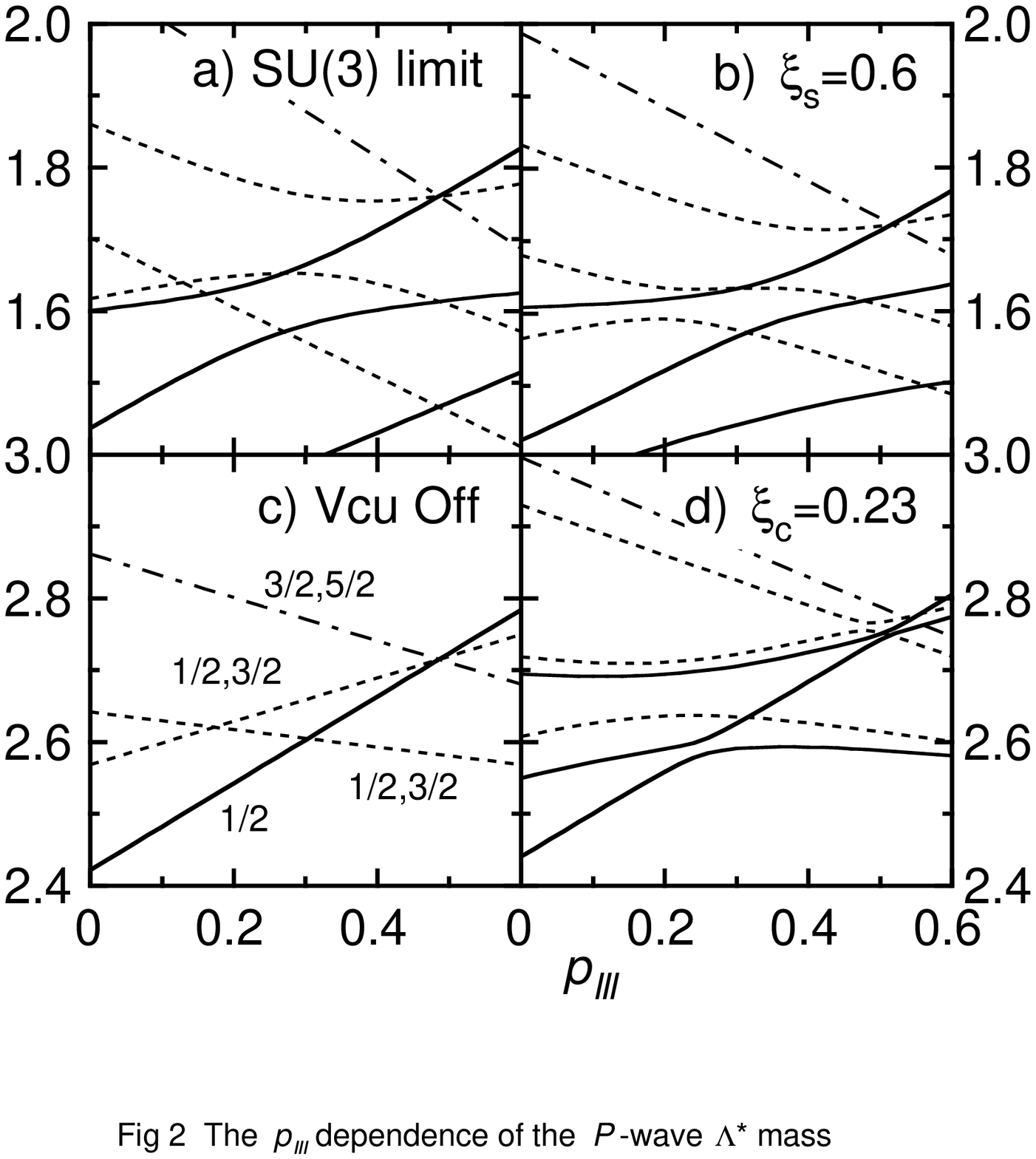}

\end{document}